\documentclass[%
reprint,
 amsmath,amssymb,
 aps,
 prl,
]{revtex4-1}

\usepackage{graphicx}
\usepackage{dcolumn}
\usepackage{bm}

\begin{document}


\title{Isomer-Dependent Fragmentation Dynamics of Inner-Shell Photoionized Difluoroiodobenzene}

\author{Utuq Ablikim $^{\textit{a,b}}$}
\author{C\'{e}dric Bomme $^{\textit{c}}$}
\author{Evgeny Savelyev $^{\textit{c}}$}
\author{Hui Xiong $^{\textit{e}}$}
\author{Rajesh Kushawaha $^{\textit{a}}$}
\author{Rebecca Boll $^{\textit{c}}$}
\author{Kasra Amini $^{\textit{d}}$}
\author{Timur Osipov $^{\textit{f}}$}
\author{David Kilcoyne $^{\textit{b}}$}
\author{Artem Rudenko $^{\textit{a}}$}
\author{Nora Berrah $^{\textit{e}}$}
\author{Daniel Rolles $^{\textit{a,c}}$}

\affiliation{$^\textit{a}$~J.R. Macdonald Laboratory, Department of Physics, Kansas State University, Manhattan, KS 66506, USA}
\affiliation{$^\textit{b}$~Advanced Light Source, Lawrence Berkeley National Laboratory, Berkeley, CA 94720, USA}
\affiliation{$^\textit{c}$~Deutsches Elektronen-Synchrotron (DESY), 22607 Hamburg, Germany}
\affiliation{$^\textit{d}$~Department of Chemistry, University of Oxford, Oxford OX1 3QZ, United Kingdom}
\affiliation{$^\textit{e}$~Department of Physics, University of Connecticut, Storrs, CT 06269, USA}
\affiliation{$^\textit{f}$~SLAC National Accelerator Laboratory, Menlo Park, CA 94025, USA}

\date{\today}

\begin{abstract}
The fragmentation dynamics of 2,6- and 3,5-difluoroiodobenzene after iodine 4$d$ inner-shell photoionization with soft X-rays are studied using coincident electron and ion momentum imaging. By analyzing the momentum correlation between iodine and fluorine cations in three-fold ion coincidence events, we can distinguish the two isomers experimentally. Classical Coulomb explosion simulations are in overall agreement with the experimentally determined fragment ion kinetic energies and momentum correlations and point toward different fragmentation mechanisms and time scales. While most three-body fragmentation channels show clear evidence for sequential fragmentation on a time scale larger than the rotational period of the fragments, the breakup into iodine and fluorine cations and a third charged co-fragment appears to occur within a few hundred femtoseconds.a time scale larger than the rotational period of the fragments, the breakup in other channels appears to occur within a few hundred femtoseconds.
\end{abstract}

\pacs{XXX}
\maketitle

\section{Introduction}
The fragmentation or \emph{Coulomb explosion} of polyatomic molecules after VUV or X-ray photoionization \cite{muramatsu_direct_2002,ueda_molecular_2005,jiang_ultrafast_2010, erk_ultrafast_2013,liekhus-schmaltz_ultrafast_2015,pitzer_absolute_2016,nagaya_femtosecond_2016}, strong-field ionization in intense laser fields \cite{hishikawa_mass-resolved_1998,sanderson_geometry_1999,legare_laser_2005,hishikawa_visualizing_2007,matsuda_visualizing_2011,pitzer_direct_2013,ibrahim_tabletop_2014}, or electron and ion impact ionization \cite{mathur_multiply_1993, moretto-capelle_fragmentation_2000,adoui_hci-induced_2001,kitamura_direct_2001,neumann_fragmentation_2010,jana_ion-induced_2011} has been investigated extensively in order to understand the dynamics of the ionization and fragmentation process as well as to study the link between the fragmentation pattern and the geometric structure of the molecules. Early experiments were mostly performed using ion time-of-flight mass spectrometry techniques such as ion-ion coincidence spectroscopy \cite{eland_dynamics_1987,eland_dynamics_1993}. The development of ion imaging techniques \cite{chandler_two-dimensional_1987, eppink_velocity_1997, hsieh_direct_1996} and, in particular, coincident ion momentum imaging \cite{dorner_cold_2000,ullrich_recoil-ion_1997,Ullrich2003,Pe2008} has significantly increased the amount of information that can be extracted from such fragmentation studies. Recently, several studies have focused on the identification of molecular isomers, i.e.~molecules with the same chemical formula but different geometric structures, from the fragmentation patterns. For example, it was demonstrated that it is possible to separate two enantiomers in a racemic mixture of small chiral molecules by measuring five-fold ion coincidences after strong-field ionization \cite{pitzer_direct_2013,pitzer_absolute_2016} or beam-foil induced Coulomb explosion \cite{herwig_imaging_2013}, while three-fold ion coincidences after inner-shell photoionization were used to identify the \emph{cis} and \emph{trans} geometric isomers of dibromoethene \cite{ablikim_identification_2016}.

Here we report on an experimental study of the fragmentation dynamics of 2,6- and 3,5-difluoroiodobenzene (C$_6$H$_3$F$_2$I; DFIB; see Fig.~\ref{fgr:setup}(b)) after iodine 4$d$ inner-shell photoionization with soft X-rays using coincident electron and ion momentum imaging. The study aims at extending coincidence momentum imaging investigations to larger and more complex molecules and, in particular, at determining if, for such complex molecules, it is possible to distinguish between the geometric structure of different isomers via coincident momentum imaging, and if the fragmentation can still be described by a simple, classical Coulomb explosion model. The choice of the particular molecules was motivated by previous work on laser-induced alignment of difluoroiodobenzene molecules \cite{viftrup_holding_2007, nevo_laser-induced_2009, ren_multipulse_2014, savelyev_jitter-correction_2017,amini_alignment_2017}, where both strong-field and soft X-ray induced Coulomb explosion were used to diagnose the degree of one- and three-dimensional molecular alignment. Since those measurements showed very distinct angular distributions of the F$^+$ fragments, we were intrigued to investigate if a coincident momentum imaging experiment that can determine the angle between the I$^+$ and F$^+$ fragment ion momenta would be able to separate the different isomers in a similar way as our previous study on dibromoethene \cite{ablikim_identification_2016}.

As we show in the following, the two isomers indeed exhibit characteristically different ion momentum correlations and fragmentation patterns that can be linked to the geometric structure of the molecules and that can be described adequately in terms of a classical Coulomb explosion model. However, the comparison of the experimental data with the Coulomb explosion simulations also reveals some distinct differences that we attribute to ultrafast charge separation across the phenyl ring as well as to a sequential breakup of the triply charged cation on a time scale of several hundred femtoseconds, which seems to occurs only in the 2,6-DFIB isomer. Other many-body fragmentation channels show clear evidence for sequential fragmentation on a time scale larger than the rotational period of the fragments.

\section{Experimental and computational details}
\label{sec:exp}
\subsection{Experimental setup}
\label{sub:exp}
\begin{figure}
\centering
  \includegraphics[height=9cm]{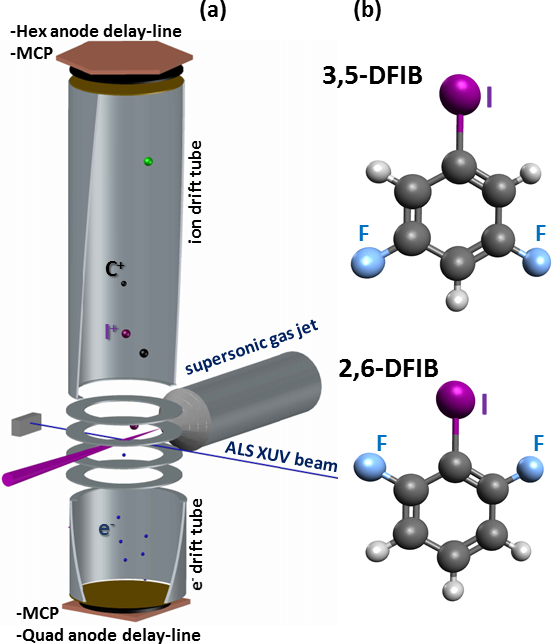}
  \caption{(a) Schematic of the experimental setup including a supersonic molecular beam and a double-sided VMI spectrometer with time- and position-sensitive delay-line detectors for coincident detection of photo-/Auger electrons and fragment ions. (b) Geometric structures of 3,5- and 2,6-difluoroiodobenzene.}
  \label{fgr:setup}
\end{figure}

The experiment was conducted at beamline 10.0.1.3 of the Advanced Light Source (ALS) at Lawrence Berkeley National Laboratory. 2,6- and 3,5-difluoroiodobenzene were commercially purchased (Sigma Aldrich, 97\% purity). Both samples are liquid at room temperature and were first outgassed in a freeze-pump-thaw cycle before introducing them into the gas phase via supersonic expansion through a 30 micron aperture using helium (backing pressure: $\approx 500$ mbar) as carrier gas.  After passing through a skimmer with a 500 micron diameter, the molecular beam was crossed by a beam of linearly polarized soft X-ray photons from the ALS (photon energy: 107 eV; bandwidth 10 meV) in the interaction center of a double-sided velocity map imaging (VMI) spectrometer. The setup, which is shown schematically in Fig.~\ref{fgr:setup}(a), was identical to the one described in a previous publication~\cite{ablikim_identification_2016} and is therefore only summarized in the following, along with a brief outline of the data analysis procedures.

In order to detect electrons and ions in coincidence and to record both position and time information for the charged fragments, which is necessary to determine their three-dimensional momentum vectors, the double-sided VMI was equipped with microchannel plate (MCP) detectors with multi-hit delay-line anodes (\emph{RoentDek DLD80} for the electrons and \emph{RoentDek HEX80} for the ions). The analog MCP and delay line signals were amplified, processed by a constant fraction discriminator (CFD), and then recorded by the data acquisition system consisting of two \emph{Roentdek TDC8HP} 8-channel multi-hit time-to-digital converters (TDC). The TDCs have a resolution of <100 ps and a multi-hit dead-time of <10 ns. They were triggered by the detection of the first electron (which could be either a photoelectron or Auger electron), which typically arrived at the detector after a flight time of approximately 5 nanoseconds. The experiment was performed during the standard ALS multi-bunch top-off mode of operation, which has an electron bunch spacing in the storage ring of 2 ns. Since this is not sufficient to link the detected photo- or Auger electron to a specific soft X-ray pulse, the time of flight of the ions was measured with respect to the arrival time of the first detected electron rather than with respect to the ALS bunch marker.

The lens voltages of the VMI spectrometer were chosen to allow for the collection of electrons up to 120 eV, singly charged ions up to 18 eV, and doubly charged ions up to 35 eV over the full solid angle. This was achieved by applying +500 and 0 V to the two inner-most extractor/repeller electrodes, +1000 and -500 V to the two additional focusing lenses, and  +/-3300 V to the two drift tubes. Since the electric field in a VMI spectrometer is not homogeneous, one cannot derive analytical formulas to reconstruct the ion momenta from the measured time of flight and hit positions of each ion. Instead, we use the \emph{SIMION 8.1} software package to simulate the expected time of flight and hit positions for ions starting in the interaction region with different kinetic energies and emission angles.
Using this procedure, the three-dimensional momentum vectors for each detected ion can be reconstructed and used to calculate the emission angles of the fragments as well as their kinetic energies. To verify the energy calibration, the kinetic energy release spectrum of N$_2$ molecules was measured, which agreed with literature values \cite{weber_k-shell_2001}.

As mentioned above, the time between two consecutive light pulses in the ALS multi-bunch operation is too short to unambiguously determine the time of flight of the electrons in order to determine their three-dimensional momenta, so only the two-dimensional projection of their momentum distribution contained in the electron hit positions is measured. The three-dimensional momentum distribution of the integrated electron image can then be reconstructed using standard VMI imaging reconstruction methods \cite{bordas_photoelectron_1996,dribinski_reconstruction_2002,manzhos_photofragment_2003,garcia_two-dimensional_2004}. For the electron spectra shown in this paper, a modified Onion Peeling method \cite{rallis_incorporating_2014} was used to invert the VMI images and reconstruct the electron spectra.

When analyzing electron-ion-ion or electron-ion-ion-ion coincidences, only those events were considered where the component-wise momentum sum of all ionic fragments falls within a narrow peak around zero with a FWHM of $15~a.u.$, which imposes momentum conservation and therefore rejects most false coincidences, i.e.~events where the fragments do not originate from the same parent molecule.
\begin{figure*}
 \centering
 \includegraphics[height=8.2cm]{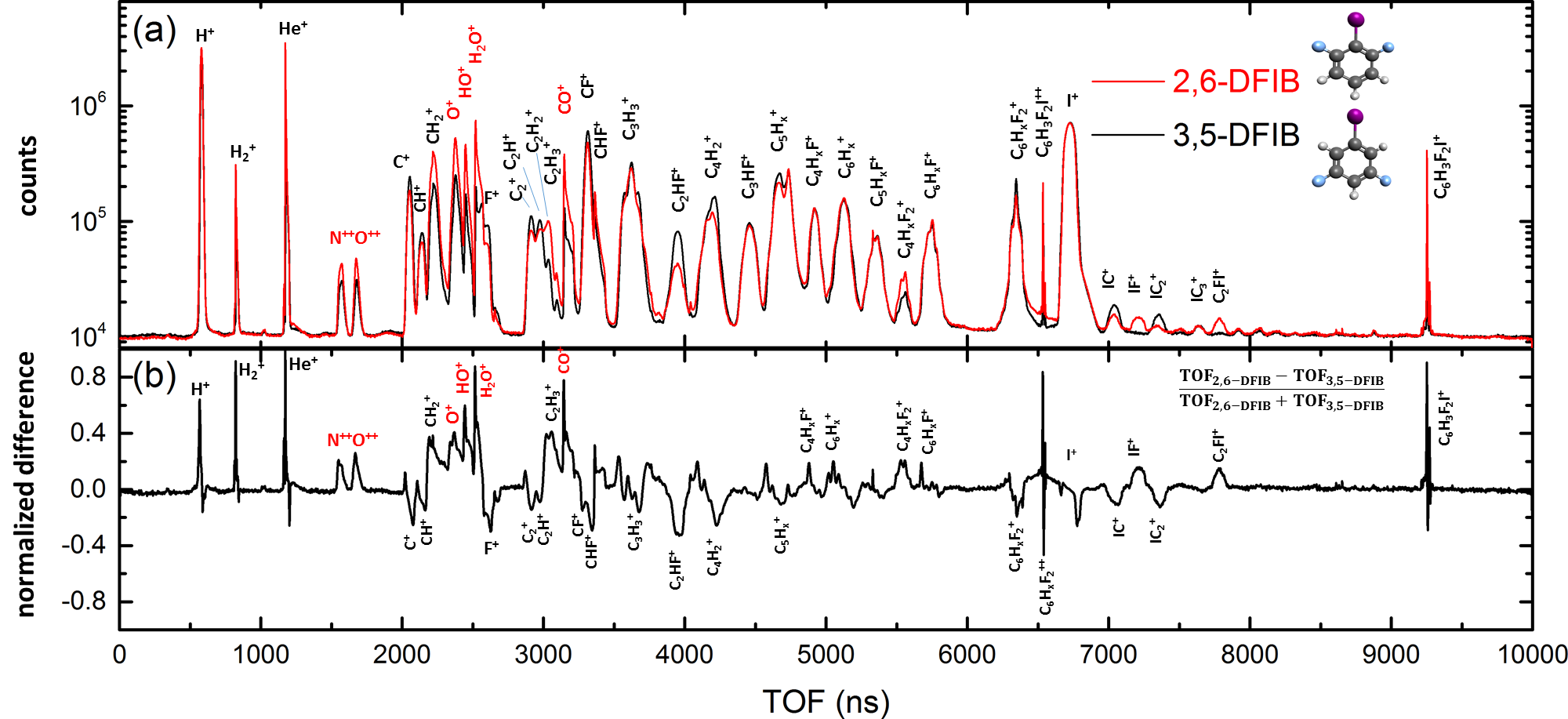}
 \caption{(a) Ion time-of-flight mass spectra generated by photoionization of 2,6- and 3,5-difluroiodobenzene at 107 eV photon energy. Peaks from residual gas are labeled in red. The spectrum of 3,5-DFIB was scaled to have the same maximum value of the I$^+$ peaks as the spectrum of 2,6-DFIB. (b) Normalized difference ($TOF_{2,6-DFIB} - TOF_{3,5-DFIB})/(TOF_{2,6-DFIB} + TOF_{3,5-DFIB})$ between the two ion mass spectra shown in the panel above.}
 \label{fgr:TOFall}
\end{figure*}

\subsection{Coulomb explosion calculations}
\label{CE}
In order to compare the experimentally determined fragment ion kinetic energies and momentum vector correlations with the expectations from a classical Coulomb explosion model, we have performed numerical simulations assuming purely Coulombic repulsion between point charges. As a starting point, we placed the charges at the center of mass of each fragment and assumed instantaneous creation of the charges followed by explosion from the equilibrium geometry of the neutral molecule, as determined by the \emph{Gaussian 09} software package \cite{gaussian09}. By numerically solving the classical equations of motions for all the fragment ions in their combined Coulomb field using a 4$^{th}$ order Runge-Kutta method, the momentum vectors and kinetic energies of all fragments were obtained for an ideal Coulomb explosion model. In order to account for possible ultrafast migration of charges inside the molecule, the calculations were repeated for other possible locations of the charges, where appropriate (see section~\ref{sec:migration}). Furthermore, in order to account for long Auger lifetimes and/or sequential bond breaking, a version of the code was implemented that allowed an increase in the charge of one of the fragments and/or the breaking of a second bond inside the molecule after a given time delay $\tau$ (see section~\ref{sec:threebody}). In that case, we simply interrupt the numerical propagation at time step $\tau$ and use the particle's positions and velocities at that moment as starting values for a new simulation with the final fragment masses and charges.

The total Coulomb energy $E_{tot}$ (in units of $eV$) of a molecules dissociating into $N$ charged fragments can also be calculated analytically as
\begin{equation}
E_{tot}~[eV]=~14.49~\sum\limits_{i\neq{j}}^{N} \frac{{q_i}{q_j}}{|{r_i}-{r_j}|},
\end{equation}
where $q_i$ and $q_j$ are the charges of the $i^{th}$ and $j^{th}$ fragment and ${|{r_i}-{r_j}|}$ is the distance between the two charges (in units of \AA) prior to the fragmentation. If no energy is stored in internal degrees of freedom, e.g.~as vibrational or rotational energy of the fragments, this formula can be used to calculate the total kinetic energy release (KER), i.e.~the sum of all fragment kinetic energies. For the case of a simple two-body fragmentation, i.e.~a break-up of the molecule into two fragments that, when combined, contain all of the atoms of the original parent molecule, the KER is partitioned, due to momentum conservation, as
\begin{equation}
E_1=\frac{m_2}{m_1+m_2} KER,~~ E_2=\frac{m_1}{m_1+m_2} KER,
\end{equation}
where $E_1$ and $E_2$ are the kinetic energies of the two fragments with masses $m_1$ and $m_2$.

\section{Results and discussion}
\label{sec:results}
\begin{figure*}
 \includegraphics[height=7.7cm]{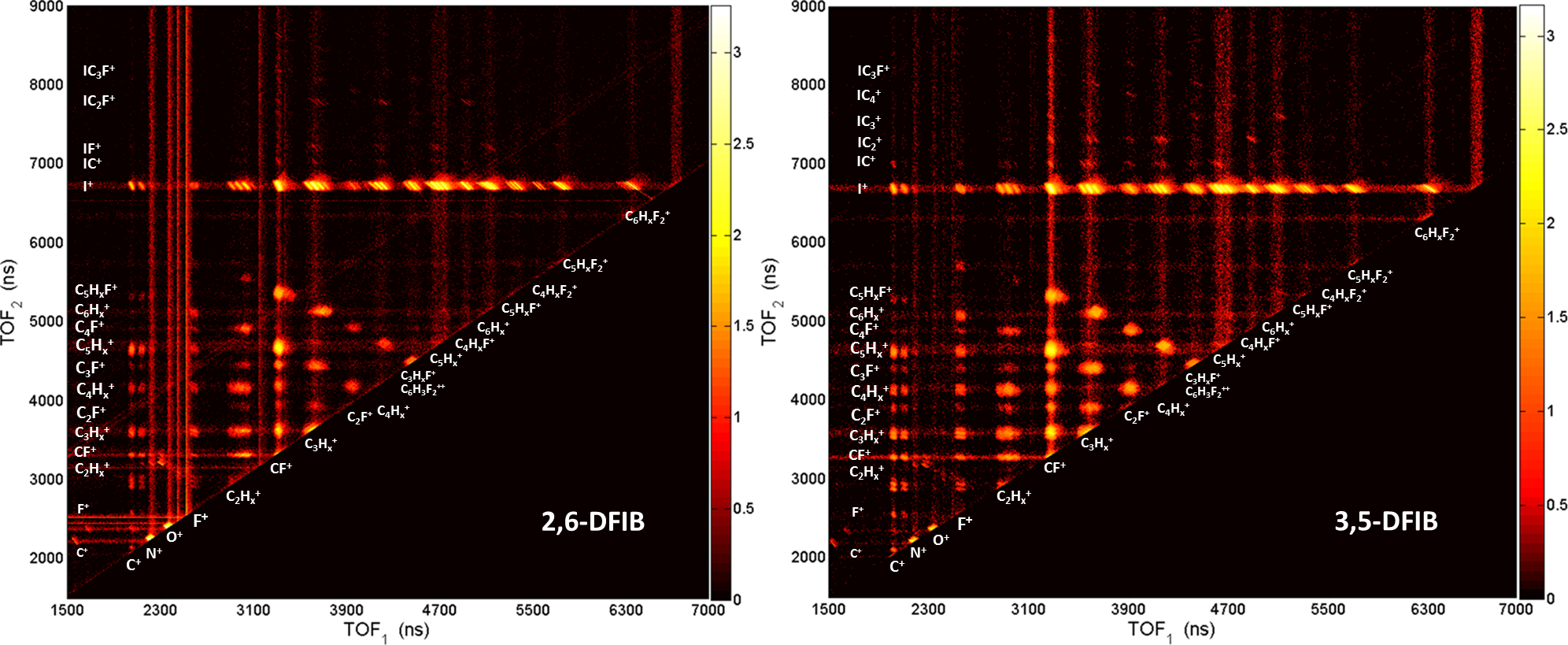}
 \caption{Photoion-photoion coincidence (PIPICO) plots for 2,6-DFIB (left) and 3,5-DFIB (right). The ion yield is shown on a logarithmic color scale.}
 \label{fgr:PIPICO26DFIB}
\end{figure*}

\begin{figure*}
 \hspace*{-0.2cm}\includegraphics[height=7.7cm]{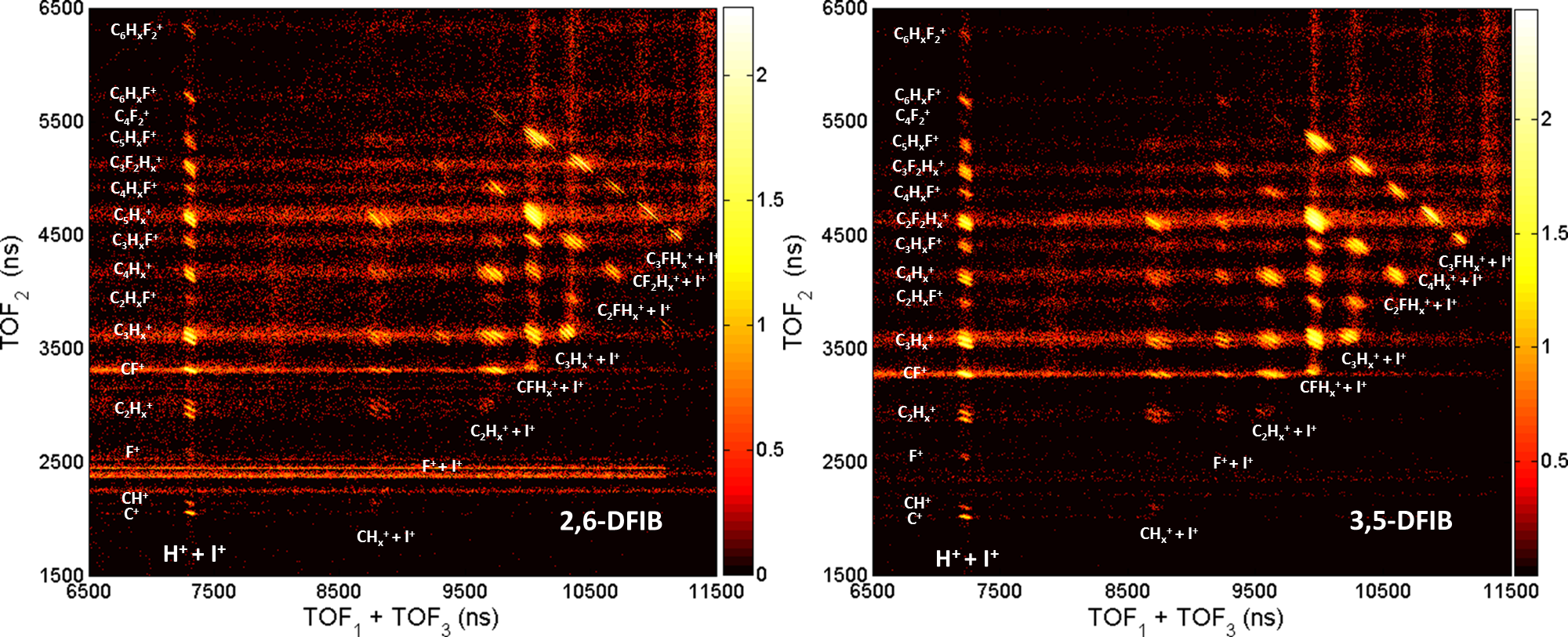}
 \caption{Triple-ion coincidence (PIPIPICO) plots for 2,6-DFIB (left) and 3,5-DFIB (right). The ion yield is shown on a logarithmic color scale.}
 \label{fgr:TriPICO5}
\end{figure*}

Fig.~\ref{fgr:TOFall}(a) shows the ion time-of-flight mass spectra of 2,6- and 3,5-difluroiodobenzene recorded at 107 eV photon energy. At this photon energy, which is approximately 50 eV above the iodine 4$d$ ionization threshold but below the iodine 4$p$ ionization thresholds in DFIB, a single photon can ionize any of the molecular valence and inner-valence shells as well as the iodine 4$d$ shell.
While valence ionization predominantly leads to singly charged final states that either remain bound or fragment into one ionic and one or several neutral fragments, emission of an I(4$d$) inner-shell photoelectron is typically followed by rapid Auger decay into doubly or triply charged cationic states. As a reference, the typical Auger lifetimes of a 4$d$-ionized Xe atom, which is electronically similar to iodine, are 6 fs for the first Auger decay and 23 fs for the second Auger step \cite{penent_multielectron_2005}, and we expect these values to be a good estimate for the order of magnitude of the lifetimes of the dominant atomic-like Auger channels in DFIB. After Auger decay, the di-cationic and tri-cationic states in DFIB generally fragment into two or three charged fragments that are emitted with relatively high kinetic energies due to the Coulomb repulsion of the positive charges (hence, this process is referred to as \emph{Coulomb explosion}). Additionally, further neutral fragments may be produced, which are not detected in this experiment. The breakup into several charged fragments can be represented in a photoion-photoion coincidence (PIPICO) plot, as shown in Fig.~\ref{fgr:PIPICO26DFIB}, where the ion yield is shown as a function of the time of flight of the first and second detected ion. The PIPICO plots for both isomers show that the molecules can break up in a large number of different channels, producing almost every charged fragment that is stoichiometrically possible. In particular, narrow diagonal lines in the PIPICO plot correspond to two-body fragmentation channel or channels where the remaining fragment(s) carry negligible momentum, while broader features correspond to breakup into three or more heavy and energetic fragments. If the molecules breaks up into three ionic fragments, one can construct a PIPIPICO (i.e.~triple ion coincidence) plot, as shown in Fig.~\ref{fgr:TriPICO5}, where the ion yield is plotted as a function of the time of flight of one of the fragments and the \emph{sum} of the times of flight of two other fragments that were detected in a given coincidence event. Again, narrow diagonal lines correspond to events, where the momenta of the three ionic fragments add to zero, while broader features correspond to breakup into more than three heavy and energetic fragments.

While the ion time-of-flight mass spectra and PIPICO/PIPIPICO plots of 2,6- and 3,5-difluroiodobenzene look rather similar at first sight, some differences, especially in the yield of F$^+$, C$_2$H$_2^+$ and fluorine containing fragments such as C$_2$HF$^+$, as well as of heavy fluorine and iodine containing fragments, such as IF$^+$ and C$2$FI$^+$, are visible upon closer inspection. This can also be seen in Fig.~\ref{fgr:TOFall}(b), where the normalized difference between the ion time-of-flight mass spectra of 2,6- and 3,5-DFIB is shown. The generation of F$^{+}$ ions from both 2,6 and 3,5-difluroiodobenzene is very rare due to the large electronegativity of fluorine, as can be seen from Fig.~\ref{fgr:TOFall}(a), but it is significantly higher in 3,5-DFIB than in 2,6-DFIB. Many of the other differences in the fragment ion yield can be explained when considering the geometry of the molecule, which favors certain fragments in one isomer as compared to the other. This is particularly evident for the C$2$FI$^+$ fragment, for example, which is only formed from 2,6-DFIB, since a C$2$FI group does not exist in the 3,5-DFIB molecule. In this context, it is interesting to point out the IF$^+$ fragment, which is only produced from 2,6-DFIB. Formation of this fragment requires the breaking of two bonds, C$-$F and C$-$I, and the formation of a new bond between the iodine and fluorine atoms. As one may intuitively expect, this bond formation only occurs in 2,6-DFIB, where iodine and fluorine are bound to neighboring carbons.

In the following, we will concentrate our discussion on the kinetic energies and momentum correlations observed in particular coincidence channels, and on the conclusions about the fragmentation dynamics that we can draw from this information.

\subsection{C$_6$H$_3$F$_2$$^{+}$ + I$^{+}$ and C$_6$H$_3$F$_2$$^{{++}}$ + I$^{+}$ two-body fragmentation channels}
\label{sec:migration}

\begin{figure}
\centering
  \includegraphics[height=6.5cm]{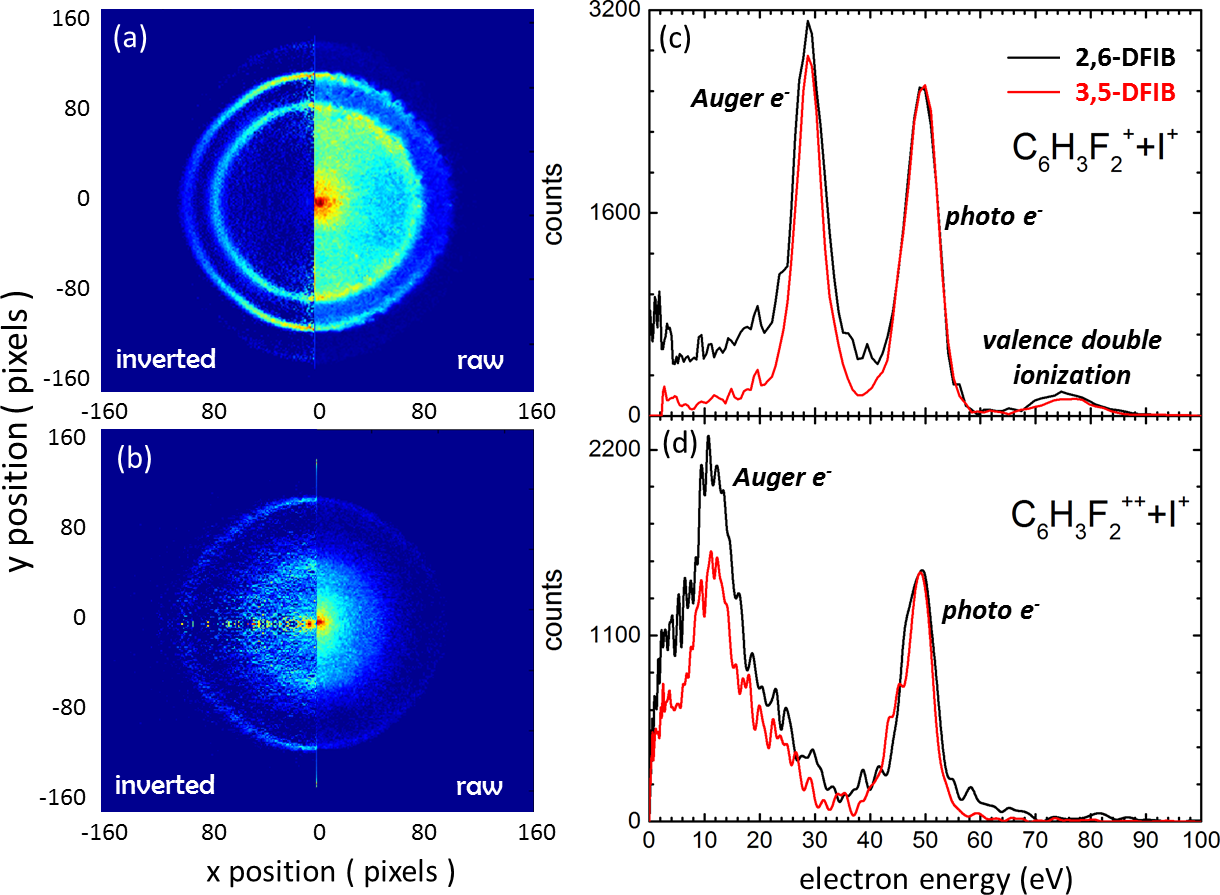}
  \caption{Velocity map electron images and kinetic energy spectra measured in coincidence with the C$_6$H$_3$F$_2$$^{+}$+I$^{+}$ (a, c) and the C$_6$H$_3$F$_2$$^{++}$+I$^{+}$ (b, d) fragment ion channels in DFIB. Panels (a) and (b) show the raw (right) and inverted (left) electron images for 2,6-DFIB, while the kinetic energy spectra in (c) and (d) are shown for both isomers.}
  \label{fgr:DFIBelectrons}
\end{figure}

As briefly mentioned in section~\ref{CE}, the conceptually easiest fragmentation channels are "complete" two-body fragmentations, where the molecule breaks into two charged fragments, which, when combined, contain all atoms that were in the original molecule. In these cases, the two fragments are emitted strictly back-to-back due to momentum conservation, and they share all of the available Coulomb energy. The strongest two-body fragmentation channel of this type is the C$_6$H$_3$F$_2$$^{+}$ + I$^{+}$ channel, which is predominantly produced by I(4$d$) inner-shell ionization followed by ultrafast Auger decay, as proven by the electron spectrum measured in coincidence with this fragmentation channel, which is shown in Fig.~\ref{fgr:DFIBelectrons}(c). The I(4$d$) photoelectrons have a kinetic energy of 50 eV, while a distinct Auger peak appears at around 29 eV kinetic energy, which is similar to the energy of the most energetic Auger lines observed after I(4$d$) ionization of CH$_3$I \cite{lindle_1984, holland_2006}. Note that there is also a smaller peak between 70 and 80 eV kinetic energy, which we attribute to valence double ionization, which also produces a doubly charged final state that can fragment into C$_6$H$_3$F$_2$$^{+}$ + I$^{+}$.

The electron spectrum for the triply charged C$_6$H$_3$F$_2$$^{++}$+I$^{+}$ final state shown in Fig.~\ref{fgr:DFIBelectrons}(d) also contains the I(4$d$) photoelectron peak, but instead of the Auger peak at around 29 eV kinetic energy, the spectrum contains a broader Auger feature with a maximum slightly above 10 eV, which is reminiscent of the lower-energetic Auger group observed in CH$_3$I \cite{lindle_1984}. Although the electron spectra are only shown for one photon energy, we have also recorded the spectra at other photon energies to confirm that the photoelectrons indeed change their kinetic energy, while the Auger electrons remain at a fixed kinetic energy, as expected.

\begin{figure}
\centering
  \includegraphics[height=6.1cm]{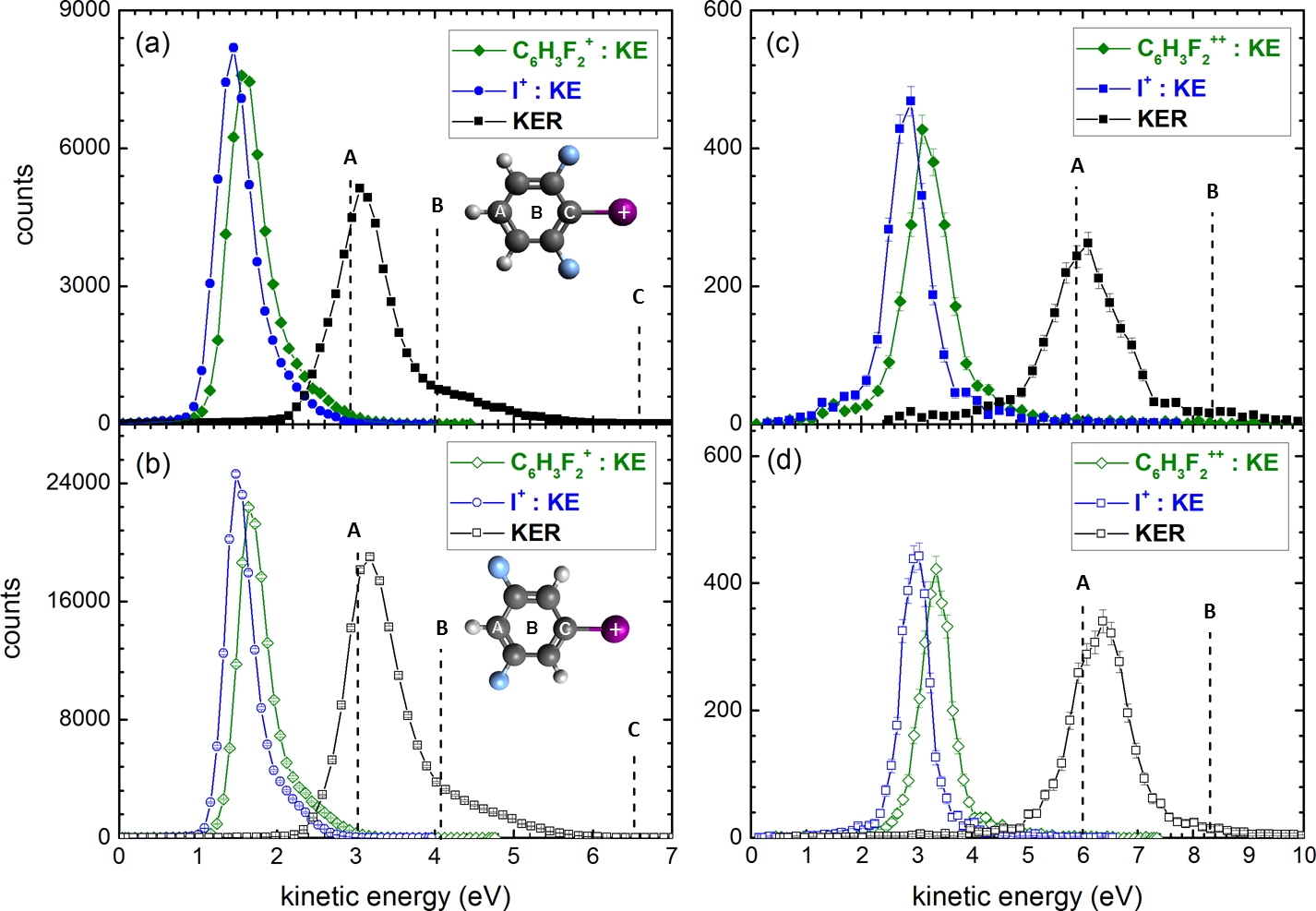}
  \caption{Kinetic energy release (black squares) of the C$_6$H$_3$F$_2$$^{+}$+I$^{+}$ (left) and C$_6$H$_3$F$_2$$^{++}$+I$^{+}$ (right) two-body fragmentation channel for 2,6-DFIB (top) and 3,5-DFIB (bottom) along with the kinetic energies of the C$_6$H$_3$F$_2$$^{+}$ or C$_6$H$_3$F$_2$$^{++}$ (green) and I$^+$ (blue) fragments. The KER values obtained from a classical Coulomb explosion simulation for three different locations of the charge(s) on the C$_6$H$_3$F$_2$$^{+}$ or C$_6$H$_3$F$_2$$^{++}$ fragments, respectively, as depicted in the insets, are shown as dashed lines. The simulated fragment ion kinetic energies for case (A) are 1.38 eV for I$^{+}$ and 1.54 eV for C$_6$H$_3$F$_2$$^{+}$.}
\label{fgr:2body4plot}
\end{figure}

The kinetic energy distributions of the C$_6$H$_3$F$_2$$^{+}$ and the I$^{+}$ fragments in the C$_6$H$_3$F$_2$$^{+}$ + I$^{+}$ coincidence channel as well as the total kinetic energy release (KER) for 2,6-DFIB and 3,5-DFIB are shown in Fig.~\ref{fgr:2body4plot}(a) and Fig.~\ref{fgr:2body4plot}(b), respectively. In both isomers, the KER is peaked at around 3.1 eV, with each fragment carrying about half of the energy since they have almost the same mass (the peaks of the experimental kinetic energy distributions are at 1.65 eV for C$_6$H$_3$F$_2$$^{+}$ and 1.45 eV for I$^{+}$). Assuming that the two charged fragments can be approximated as point charges and that the molecule breaks up on a purely Coulombic potential energy curve after both charges are created, we can calculate the Coulomb energy of the system for different locations of the two charges, as described in section~\ref{CE}. The dashed lines in Fig.~\ref{fgr:2body4plot} show the value of this Coulomb energy if one of the two charges is localized on the iodine fragment, while the other one is located at three different positions on the phenyl ring: (A) on the carbon atom furthest away to the iodine, (B) at the center of the ring, and (C) on the carbon atom closest to the iodine. Case (A) agrees almost perfectly with the maximum of the measured KER distribution, case (B) lies in the high energy "shoulder" of the KER distribution, while case (C) clearly overestimate the energy significantly. From this, we can conclude that either (i) the fragmentation does not occur along a Coulombic potential curve and a significant fraction of the Coulomb energy is transformed into internal energy, e.g.~in electronic, vibrational or rotational excitations, (ii) the C$-$I bond has stretched significantly before the second charge was created, or (iii) the second charge has localized at the far end of the phenyl ring before the Coulomb explosion occured.

Since we cannot distinguish between these possibilities without detailed quantum chemistry calculations, we turn to another two-body fragmentation channel to obtain further information. Fig.~\ref{fgr:2body4plot}(c) and (d) show the measured fragment ion kinetic energy distributions and KER for the C$_6$H$_3$F$_2$$^{++}$ + I$^{+}$ channel in both 2,6- and 3,5-DFIB, compared to the calculated Coulomb energies for the three scenarios described above. Again, the situation where both charges on the C$_6$H$_3$F$_2$$^{++}$ fragment are located at the far end of the phenyl ring gives almost perfect agreement with the experimental data. Since it is unlikely that the amount of internal energy in the molecular fragment, which would have to be 6 eV to explain the difference, would have increased so drastically in this case as compared to the doubly charged fragmentation channel, we conclude that ultrafast charge localization is the most likely scenario: After photoionization removes an I(4$d$) electron, the inner-shell vacancy in the iodine atom is filled by a valence or inner-valence electron via an Auger process that ejects a second and sometimes a third valence electron. This leaves the system with two or three holes in the valence shell. Charge migration along the phenyl ring, driven by the Coulomb repulsion between the holes, could lead to a situation where the holes are located at opposite ends of the molecule before the molecule fragments.

Similar ultrafast charge migration after inner-shell ionization of a benzene compound was recently suggested in a theoretical study of nitrosobenzene molecules \cite{kuleff_core_2016}. In this study, the authors investigated charge migration in the valence shell driven solely by electron correlation and electron relaxation. The calculations show that in core-ionized nitrosobenzene, charge migration occurs within less than 1 femtosecond and, in particular, even faster than the Auger decay. From the present experimental data, we do not have direct evidence for such a charge migration effect in DFIB nor can we draw any conclusions about the mechanism for charge localization, but we note that this process would explain the experimentally observed fragment energies.

The differences in the yield of F$^{+}$ ions seen in Fig.~\ref{fgr:TOFall}, which we pointed out earlier, may further support this hypothesis: If charge migration leads to a positive charge at the end of the ring opposite to the iodine, i.e.~close to the fluorine atoms in 3,5-DFIB, a lack of electrons in the vicinity of the fluorines might make it more likely to produce F$^+$ ions than in the case of 2,6-DFIB, where the positive charge on the phenyl ring is located further away from the fluorine atoms.

\subsection{Sequential breaking of C$-$I and C$-$C bonds}
\label{sec:sequential}

While the majority of DFIB molecules are in a doubly ionized final state after I(4$d$) inner-shell ionization and subsequent Auger decay, a significant fraction
of the molecules end up in a triply charged final state, as demonstrated by Fig.~\ref{fgr:TriPICO5}. This can happen via direct double ionization, most likely via a \emph{shake-off} process, in the first ionization step followed by a single Auger process, or via emission of a single photoelectron followed by emission of two Auger electrons, either simultaneously (\emph{double-Auger}) or sequentially (\emph{Auger cascade}) \cite{Viefhaus2005,lablanquie_multielectron_2007}. For the triply charged C$_6$H$_3$F$_2$$^{++}$+I$^{+}$ final state, the electron spectrum in Fig.~\ref{fgr:DFIBelectrons}(d) clearly shows that this state is reached via single photoelectron emission, since direct double photoionization would not yield a well-defined photoline at 50 eV kinetic energy. This first step is followed, most likely, by a sequential Auger cascade, since "double-Auger" emission would also produce a more continuous electron kinetic energy distribution than what is observed here.
\begin{figure}
\centering
\includegraphics[height=9cm]{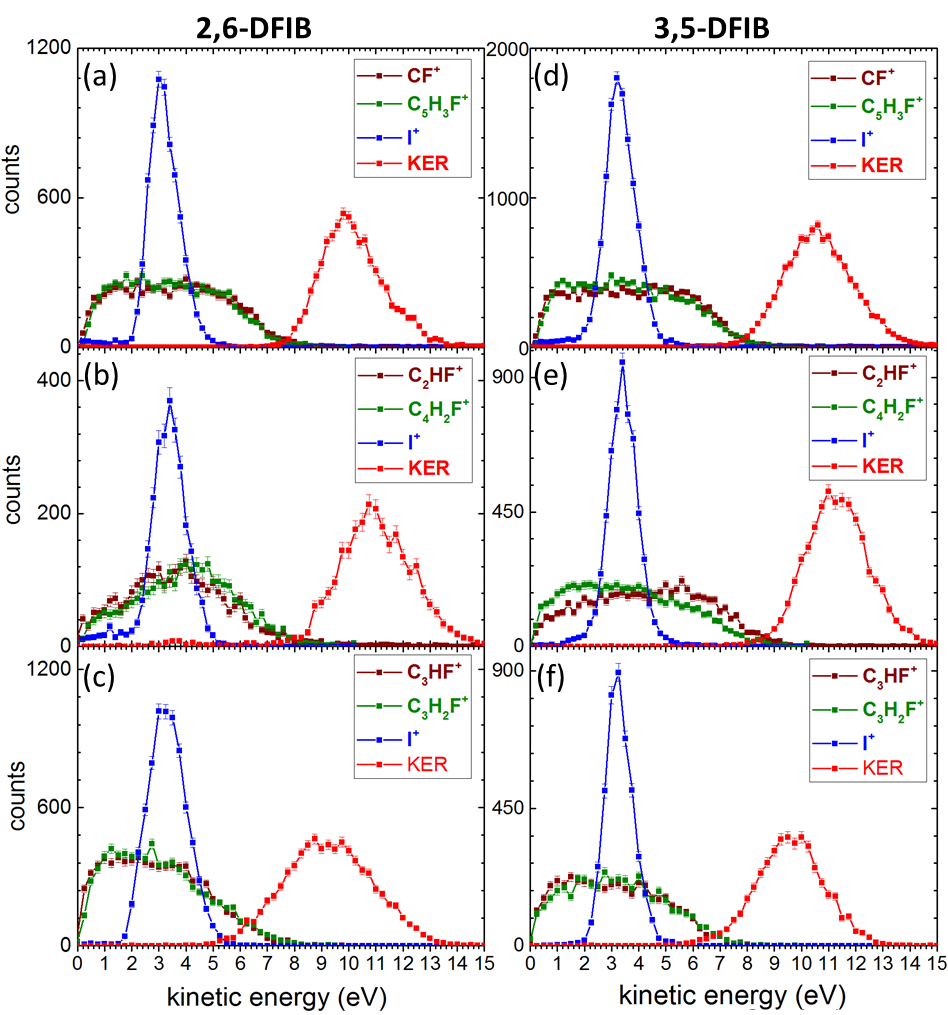}
  \caption{Kinetic energies of individual ionic fragments and total kinetic energy release for the CF$^+$+C$_5$H$_3$F$^+$+I$^+$ (top), C$_2$HF$^+$+C$_4$H$_2$F$^+$+I$^+$ (middle), and C$_3$HF$^+$+C$_3$H$_2$F$^+$+I$^+$ (bottom) channels in 2,6-DFIB (left) and 3,5-DFIB (right).}
  \label{fgr:SeqKERs}
\end{figure}

Since the triply charged DFIB parent ion is not stable, it breaks up in two or three charged fragments and, possibly, further neutral fragments. The events where the molecule breaks into three charged fragments are shown in the triple-ion coincidence maps in Fig.~\ref{fgr:TriPICO5}. The strongest contributions are from fragmentation channels where an I$^{+}$ ion and two fragments from the phenyl ring are produced. Here we concentrate on three exemplary triple coincidence channels, namely CF$^+$+C$_5$H$_3$F$^+$+I$^+$, C$_2$HF$^+$+C$_4$H$_2$F$^+$+I$^+$, and C$_3$HF$^+$+C$_3$H$_2$F$^+$+I$^+$. Their coincident electron spectra are qualitatively similar to those of the C$_6$H$_3$F$_2$$^{++}$+I$^{+}$ final state shown in Fig.~\ref{fgr:DFIBelectrons}(d), but the statistics and kinetic energy resolution of our data are not sufficient to observe possible subtle differences in the Auger electron spectrum.

After obtaining the three-dimensional momenta of all ionic fragments in these coincidence channels, the individual fragment ion kinetic energies and the KERs are calculated and are shown in Fig.~\ref{fgr:SeqKERs}. For all three fragmentation channels, the kinetic energies are very similar in both isomers, 2,6- and 3,5-DFIB. Furthermore, it is interesting to note that the narrow kinetic energy distributions of the iodine ions are almost identical to those in C$_6$H$_3$F$_2$$^{++}$+I$^+$ two body Coulomb explosion channel.
\begin{figure}[ht]
\centering
\includegraphics[height=11cm]{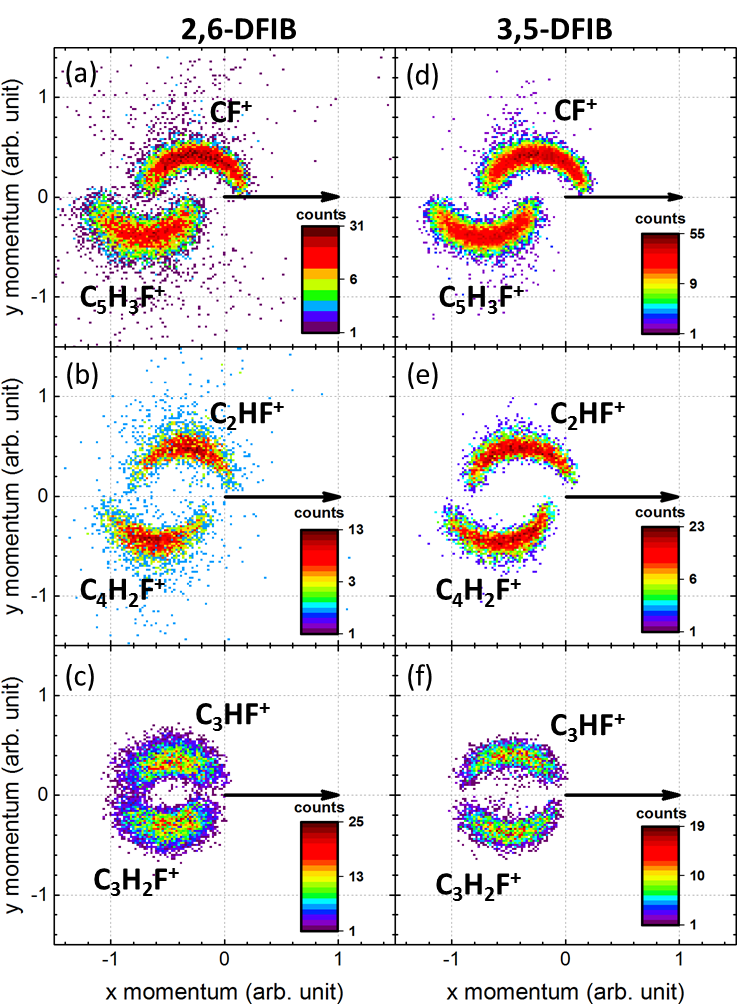}
  \caption{Newton plots of the CF$^+$+C$_5$H$_3$F$^+$+I$^+$, C$_2$HF$^+$+C$_4$H$_2$F$^+$+I$^+$, and C$_3$HF$^+$+C$_3$H$_2$F$^+$+I$^+$ coincidence channels in 2,6-DFIB (left) and 3,5-DFIB (right). The momentum of the I$^+$ fragment is plotted as a unit vector (black arrow), while the momenta of the two other ionic fragments relative to the I$^+$ momentum are plotted in the upper and lower half, respectively. The shift between the upper and lower semi-circular structures in the asymmetric break-up channels is caused by the large mass difference between the fragments, which results in very unequal sharing of the center-of-mass momentum from the first fragmentation step.}
  \label{fgr:NewtonSeq}
\end{figure}

\begin{figure}
\centering
\includegraphics[height=9.2cm]{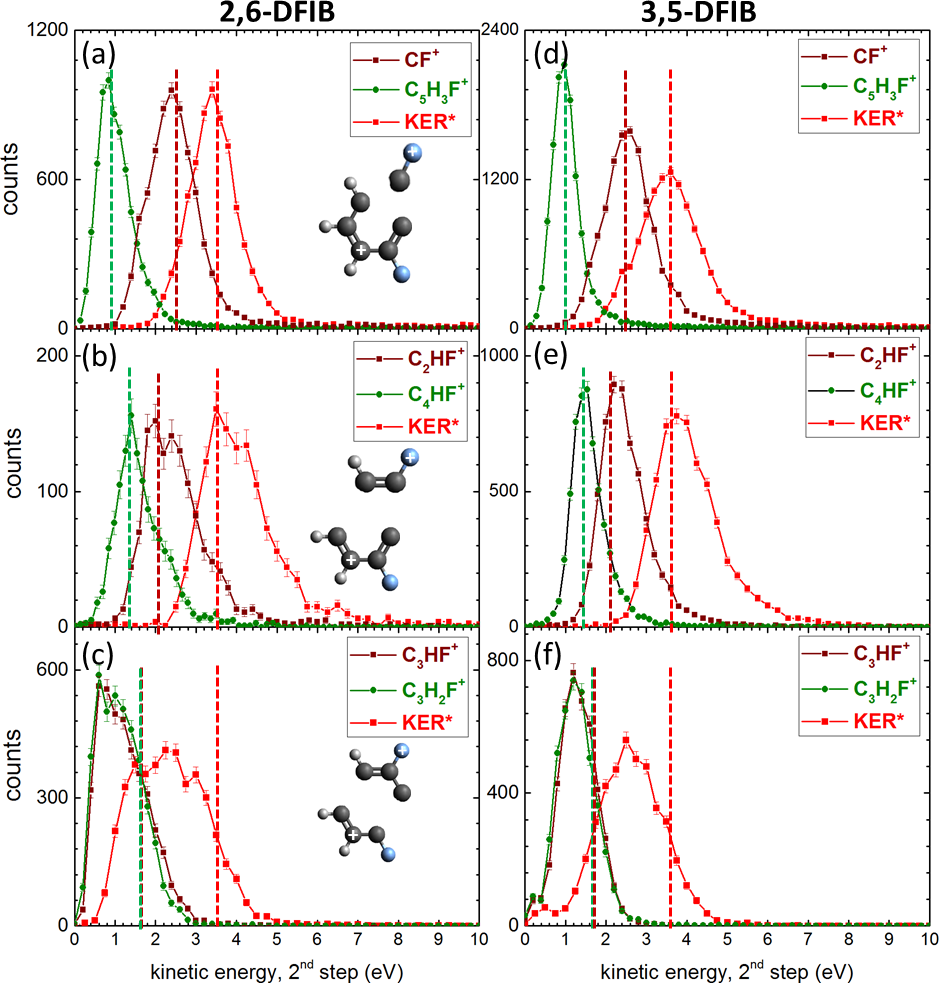}
  \caption{Kinetic energies of individual fragments and total kinetic energy release for the second fragmentation step in the three-body fragmentation channels shown in Fig.~7 and 8 for 2,6-DFIB (left) and 3,5-DFIB (right). The kinetic energies obtained from a classical Coulomb explosion simulation are shown as dashed lines.}
  \label{fgr:KER2}
\end{figure}

In order to gain further insight into the fragmentation mechanism leading to these three-body channels, Newton plots are shown in Fig.~\ref{fgr:NewtonSeq}, where the momenta of the two carbon-containing fragments are plotted with respect to the momentum of the iodine ion, which is represented by a black arrow. The curved, semi-circular structures that appear in these Newton plots are a strong indication for a sequential fragmentation \cite{neumann_fragmentation_2010}, with a delay between the breaking of the C$-$I bond and the subsequent breaking of the C$-$C bonds longer than the rotational period of the C$_6$H$_3$F$_2$$^{++}$ fragment, which is on the order of 100 $ps$ in the rotational ground state. We can thus hypothesize that the process leading to these three-body channels proceeds as follows: Inner-shell photoionization followed by emission of two Auger electrons leaves the molecule in a triply charged state, which undergoes Coulomb explosion into C$_6$H$_3$F$_2$$^{++}$+I$^+$, leading to a singly charged iodine ion with about 3 eV final kinetic energy and a metastable C$_6$H$_3$F$_2$$^{++}$ di-cation with about 3.5 eV kinetic energy, both repelled in opposite directions. At the same time, the C$_6$H$_3$F$_2$$^{++}$ fragment seems to have received some angular momentum during the C$-$I bond breaking (e.g.~as the result of C$-$I bending vibrations), resulting in a rotation around its center of mass. After a delay longer than its rotational period, the metastable C$_6$H$_3$F$_2$$^{++}$ di-cation breaks up into two singly charged fragments, each containing a fluorine atom and different numbers of carbons atoms. At that time, the distance to the iodine ion is large enough that the Coulomb force on the I$^+$ ion is negligible, as demonstrated by the narrow I$^+$ kinetic energy distribution, which is independent of the secondary fragmentation. Under this assumption, we can retrieve the kinetics of the second-step fragmentation by subtracting the center-of-mass velocity of the C$_6$H$_3$F$_2$$^{++}$ di-cation, which can be calculated from the measured I$^+$ momentum because of momentum conservation, from each of the other fragment velocities, thus retrieving the kinetic energy spectrum of the second Coulomb explosion step, which is shown in Fig.~\ref{fgr:KER2}. Again, our classical model simulation, shown as dashed lines, are in good agreement with the experimental data, suggesting that the second-step decay also occurs along Coulombic potential curves. To obtain the best match with the experimental kinetic energies, we placed the two charges in the second Coulomb explosion step on the fluorine atom in one of the fragments and on the carbon atom in the second fragment that is furthest away from the first fluorine atom. This yields very good agreement in the two asymmetric fragmentation channels but overestimates the kinetic energies in the symmetric C$_3$HF$^+$+C$_3$H$_2$F$^+$+I$^+$ fragmentation, suggesting an intermediate geometry where the two charges are even further apart in that case, possibly due to a deformed geometry of the metastable di-cation.

\subsection{Identification of molecular isomers via fragment ion momentum correlations in three-body fragmentation channels}
\label{sec:threebody}
\begin{figure}
\centering
  \includegraphics[height=4.0cm]{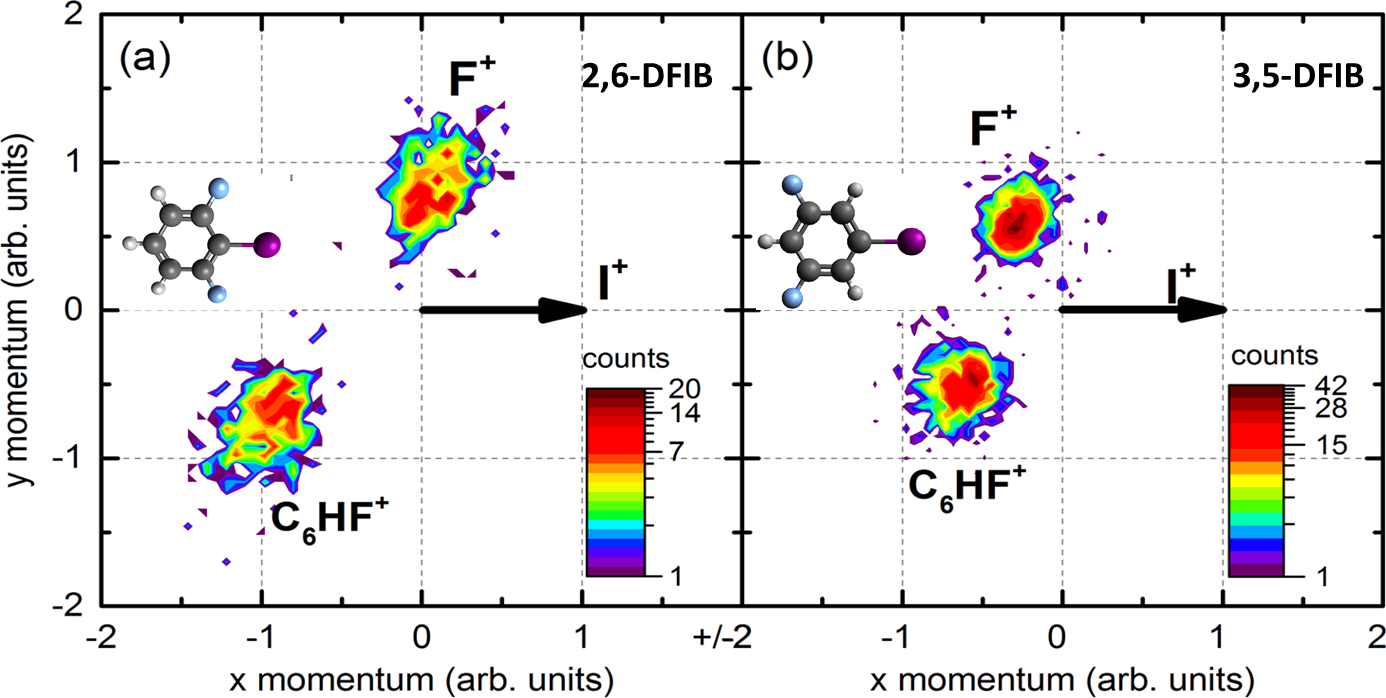}
  \caption{Newton plots of the F$^{+}$ + C$_6$HF$^{+}$ + I$^{+}$ fragmentation channel for (a) 2,6-DFIB and (b) 3,5-DFIB. The momentum vectors of F$^{+}$ and C$_6$HF$^{+}$ are normalized to the size of the momentum vector of I$^{+}$.}
  \label{fgr:newton}
\end{figure}
\begin{figure}
\centering
\includegraphics[height=6.2cm]{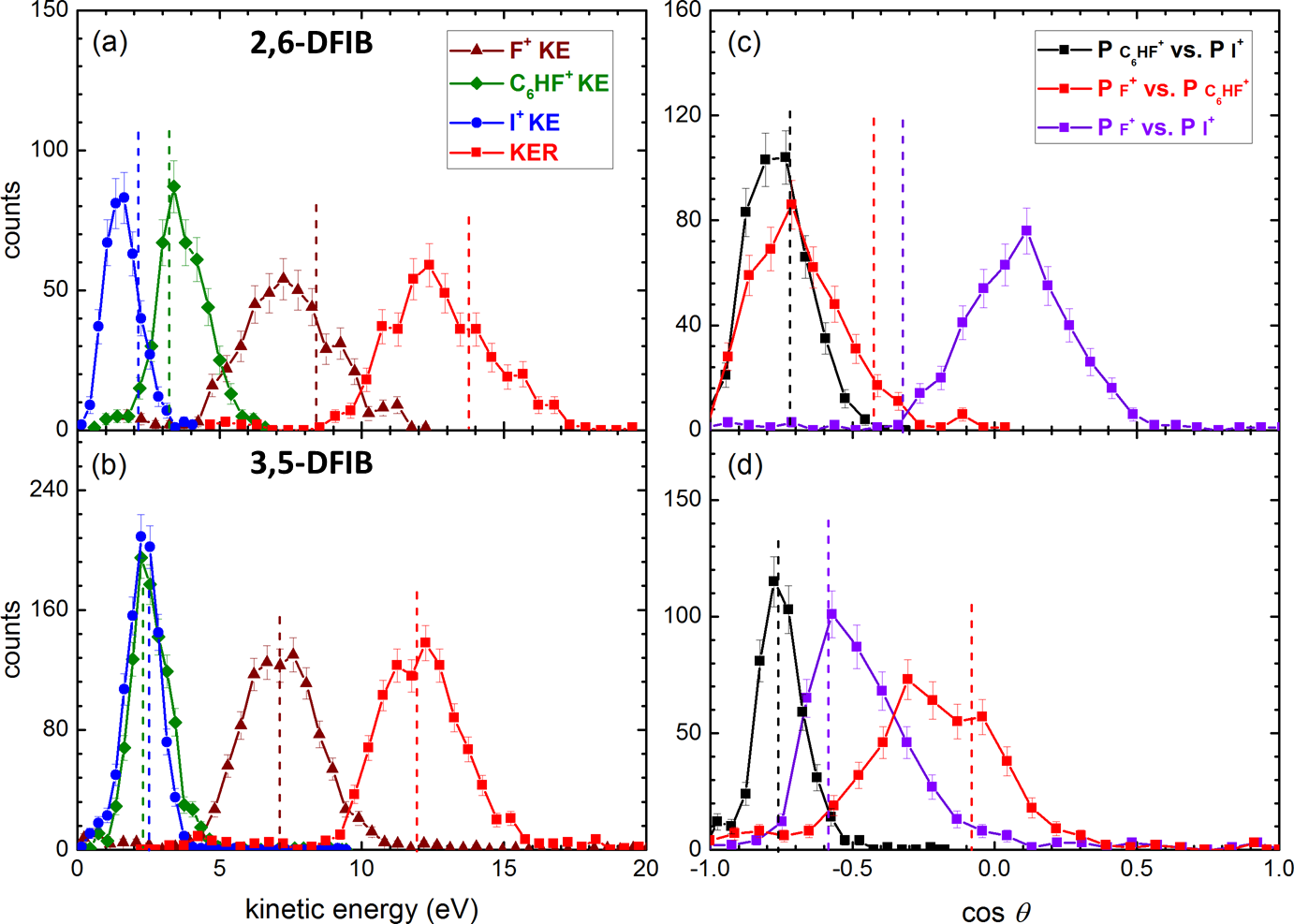}
  \caption{Total kinetic energy release, kinetic energies of the individual ionic fragments, and angle $\theta$ between the momentum vectors of the F$^+$, I$^+$, and C$_6$HF$^{+}$ fragments for the F$^{+}$ + C$_6$HF$^{+}$ + I$^{+}$ fragmentation channel in 2,6-DFIB (top) and 3,5-DFIB (bottom). The kinetic energies and angles obtained from a classical Coulomb explosion simulation are shown as dashed lines.}
  \label{fgr:FpC6FHpIp4plot}
\end{figure}

As we have shown previously for the case of dibromoethene \cite{ablikim_identification_2016}, the momentum correlations in certain three-body fragmentation channels can be used to identify geometric isomers, in that case by determining the angle between the momentum vectors of the two Br$^+$ ions that are emitted in coincidence with a C$_2$H$_2^+$ fragment. In the case of DFIB, one might expect that the angle between I$^+$ and F$^+$ fragments could be used to distinguish between 2,6- and 3,5-DFIB, if the fragmentation happens fast enough to preserve the angular correlation between these two fragments. We first concentrate on the F$^{+}$ + C$_6$HF$^{+}$ + I$^{+}$ fragmentation channel, in which all the heavy atoms are accounted for in the ionic fragments, and only two hydrogen atoms are missing. They were most likely emitted as neutral fragments, since the momentum sum of the three ionic fragments is very narrow around zero (FWHM=$\pm$7.5 a.u). Fig.~\ref{fgr:newton} shows the Newton plots for this fragmentation channel in both 2,6- and 3,5-DFIB. The first observation from these Newton plots, where the momenta of two fragments (F$^+$ and C$_6$HF$^{+}$) are plotted in the frame of the momentum of the third fragment (I$^+$), is that they show well-defined peaks rather than smeared out circular structures, suggesting that both bond breaks between the charged fragments happen on a time scale faster than the molecular rotation. Furthermore, there is a clear difference in  the fragmentation patterns of the two isomers, with smaller relative momenta of the F$^+$ and C$_6$HF$^{+}$ fragments in the case of 3,5-DFIB and a larger angle between I$^+$ and F$^+$ fragments as compared to 2,6-DFIB. The difference in the fragmentation patterns for the two isomers is also very apparent in Fig.~\ref{fgr:FpC6FHpIp4plot}, where the KER and the fragment ion kinetic energies for this channel are shown for both isomers, along with the angle $\theta$ between the momentum vectors of the F$^+$, I$^+$, and C$_6$HF$^{+}$ fragments detected in coincidence. The KER and F$^+$ kinetic energies are rather similar in 2,6- and 3,5-DFIB, with the main difference being a lower I$^+$ and higher C$_6$HF$^{+}$ kinetic energy in 2,6-DFIB as compare to 3,5-DFIB, where both fragments have almost identical kinetic energies. The angles show large differences between the two isomers, with the angle between F$^+$ and I$^+$ fragments peaking around around $84\,^{\circ}$~($cos~\theta=0.1$) for 2,6-DFIB as opposed to $120\,^{\circ}$~($cos~\theta=-0.6$) for 3,5-DFIB.

While these plots show that the momentum correlation between the F$^+$ and I$^+$ fragments can indeed be used to separate and identify the two isomers, the experimentally observed angles are surprising for 2,6-DFIB, where one may have naively expected a smaller angle between F$^+$ and I$^+$ since the angle between the F and I atoms in the equilibrium geometry of the neutral 2,6-DFIB molecule is $61\,^{\circ}$. The Coulomb explosion simulation for the three-body fragmentation shows that this naive expectation is not justified, since the charged fragments repel each other in a way that the angles between the detected ion momenta are not necessarily equal or even close to the bond angles in the molecule. While the Coulomb explosion simulations, shown as dashed lines in Fig.~\ref{fgr:FpC6FHpIp4plot}, are in good agreement with the experimentally observed kinetic energies and angles in 3,5-DFIB, they do not reproduce the observed angles for 2,6-DFIB. In this simulation, one charge is placed at the position of the iodine atom, the second at the position of the fluorine atom, and the third one in the center of the ring. Both the C$-$I and the C$-$F bond are assumed to break simultaneously, a scenario commonly referred to as \emph{concerted fragmentation}. For 2,6-DFIB, concerted fragmentation for any charge configuration yields angles between the fragments that do not match the experimentally determined angles at all. We have tried various other possible positions of the charge on the C$_6$HF$^{+}$ and found that none of them can reproduce the experimentally observed energies and/or angles. In particular, they all yield too large of an angle between the F$^{+}$ and I$^{+}$ fragments. Interestingly, for some charge configurations, concerted fragmentation of 2,6-DFIB can even lead to F$^{+}$-I$^{+}$ angles that are very similar to those observed in 3,5-DFIB, suggesting that the seemingly "obvious" link between the molecular geometry and the fragment angle correlations should be considered with caution and on a case-by-case basis, rather than as a general rule.

One scenario that would lead to a smaller angle between F$^{+}$ and I$^{+}$ fragments would be a step-wise ionization and/or fragmentation, where the I$-$C bond is broken first, e.g.~after the first Auger decay, and the the remaining C$_6$H$_x$F$_2$$^{+}$ remains in an excited state that decays, via a second Auger decay, after a few hundred femtoseconds, when the distance to the iodine has already increased considerably due to the first Coulomb explosion step. A further indication for such as delay of the second-step Auger decay is the kinetic energy of the I$^{+}$ fragment, which is significantly lower than any concerted fragmentation scenario would allow.

\begin{figure} [ht]
\centering
  \includegraphics[height=11.3cm]{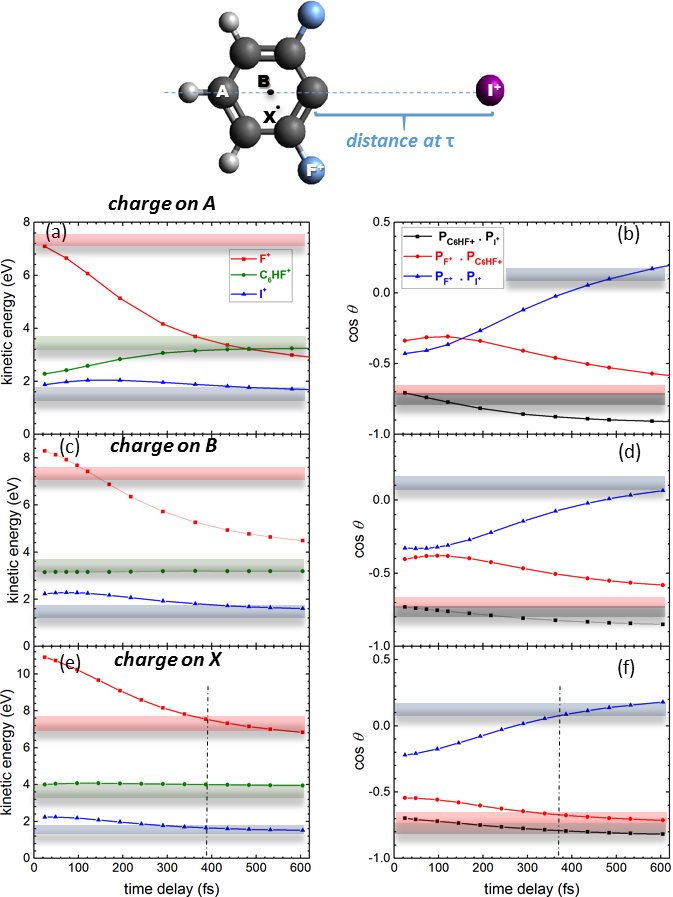}
  \caption{Fragment ion kinetic energies and angles between the momentum vectors obtained from the Coulomb explosion simulation of a two-step fragmentation of 2,6-DFIB with a variable time delay $\tau$ between the two fragmentation steps (see text) and for different locations of the charge on the C$_6$HF$^+$ fragment, as shown in the sketch above. In (a) and (b), one of the charges in the second-step Coulomb explosion is placed on the carbon furthest away from the iodine (labeled A), in (c) and (d), it is placed in the center of the phenyl ring (labeled B), and in (e) and (f), it is placed close to the fluorine atom at the position labeled X. The second charge is always on the fluorine atom. The experimentally observed kinetic energies and angles are shown as shaded areas.
The vertical black dashed lines in the bottom panels mark the delay $\tau$, at which the simulations agree best with the experimental values.}
  \label{fgr:two_step_sim_Y}
\end{figure}
As described in section 2.1, we can model a delayed ionization and fragmentation by introducing a time $\tau$, after which the charge of a specific fragment is increased and/or a specific bond is broken. Fig.~\ref{fgr:two_step_sim_Y} shows the result of these calculations for step-wise fragmentation of 2,6-DFIB for the case that the two charges, in the first step, are located at the position of the iodine atom and the carbon atom that is furthest away from it, and for two different locations of the charges in the second step, as a function of the delay $\tau$ between the two steps. When assuming that the charge on the C$_6$HF$_2$$^{+}$ fragment in the second step is located in the vicinity of the F$^{+}$ fragment, we can reproduce all of the experimentally observed kinetic energies and angles, within reasonable accuracy, at a delay $\tau$ around 400 fs, as shown in Fig.~\ref{fgr:two_step_sim_Y}(e) and (f). Note that this delay is still significantly shorter than the rotational period of the C$_6$HF$_2$$^{+}$ fragment, such that no "smearing out" of the angles can be seen in the Newton plot in Fig.~\ref{fgr:newton}(a). Any other scenario (including many more that we have tried but that are not shown here) yields significant deviations in at least one observable. Without having a direct proof for this hypothesis beyond the agreement between the Coulomb explosion simulation and the experimental data, we tentatively explain our observation as follows: After the initial Auger decay following the creation of a I($4d$) vacancy, there is a certain probability that the molecule fragments into an I$^{+}$ and an electronically excited C$_6$H$_x$F$_2$$^{+*}$ fragment. If the electronic energy in the C$_6$H$_x$F$_2$$^{+*}$ is sufficient, e.g.~if it has an inner-valence hole, this fragment can decay further into a multitude of tri-cationic channels that can be seen in Fig.~\ref{fgr:TriPICO5}. Most of these secondary Auger decays will occur much faster than the $\approx$400 fs lifetime that we derive from our simulation, leading, e.g., to the three-body fragmentation channels discussed in section 3.2. However, since the fragmentation into F$^{+}$ + C$_6$HF$^{+}$ + I$^{+}$ is a rather weak channel, it is conceivable that it only occurs after an initial Auger decay into a rather long-lived state with an inner-valence vacancy with a lifetime on the order of $\approx$ 400 fs. Furthermore, since fluorine has a very high electronegativity, it is very unlikely to dissociate into a F$^{+}$ fragment, unless the inner-valence vacancy in the C$_6$H$_x$F$_2$$^{+}$ is located in an orbital that has significant overlap with one of the near-atomic fluorine orbitals. Even without further calculations, it is therefore conceivable that the different geometry of 2,6- and 3,5-DFIB and, in particular, the different location of the fluorine atoms with respect to the iodine atom, may lead to significantly different lifetimes of the intermediate states that lead to this particular fragmentation channel.

Of course, the classical Coulomb explosion model is unable to test or predict any of these detailed electronic dynamics, but it seems to be clear that a step-wise fragmentation model needs to be considered in order to obtain satisfying agreement with the experimental data for 2,6-DFIB. We further note that there is not only an ambiguity in the exact positioning of the charge in the model, but also in the geometry of the intermediate state. This leads to an uncertainty of the delay, for which we can achieve satisfactory agreement of the simulated kinetic energies and angles with the experimental data. We have not attempted to perform a multi-parameter least-square fitting procedure to obtain a more accurate number for the delay $\tau$, since the classical Coulomb explosion model is not suitable to draw such precise and quantitative conclusions. Nevertheless, it yields, at least, an estimate for the lifetime of the second-step Auger process, if the assumption of a two-step Auger process is correct.
\begin{figure}[ht]
\centering
  \includegraphics[height=5.8cm]{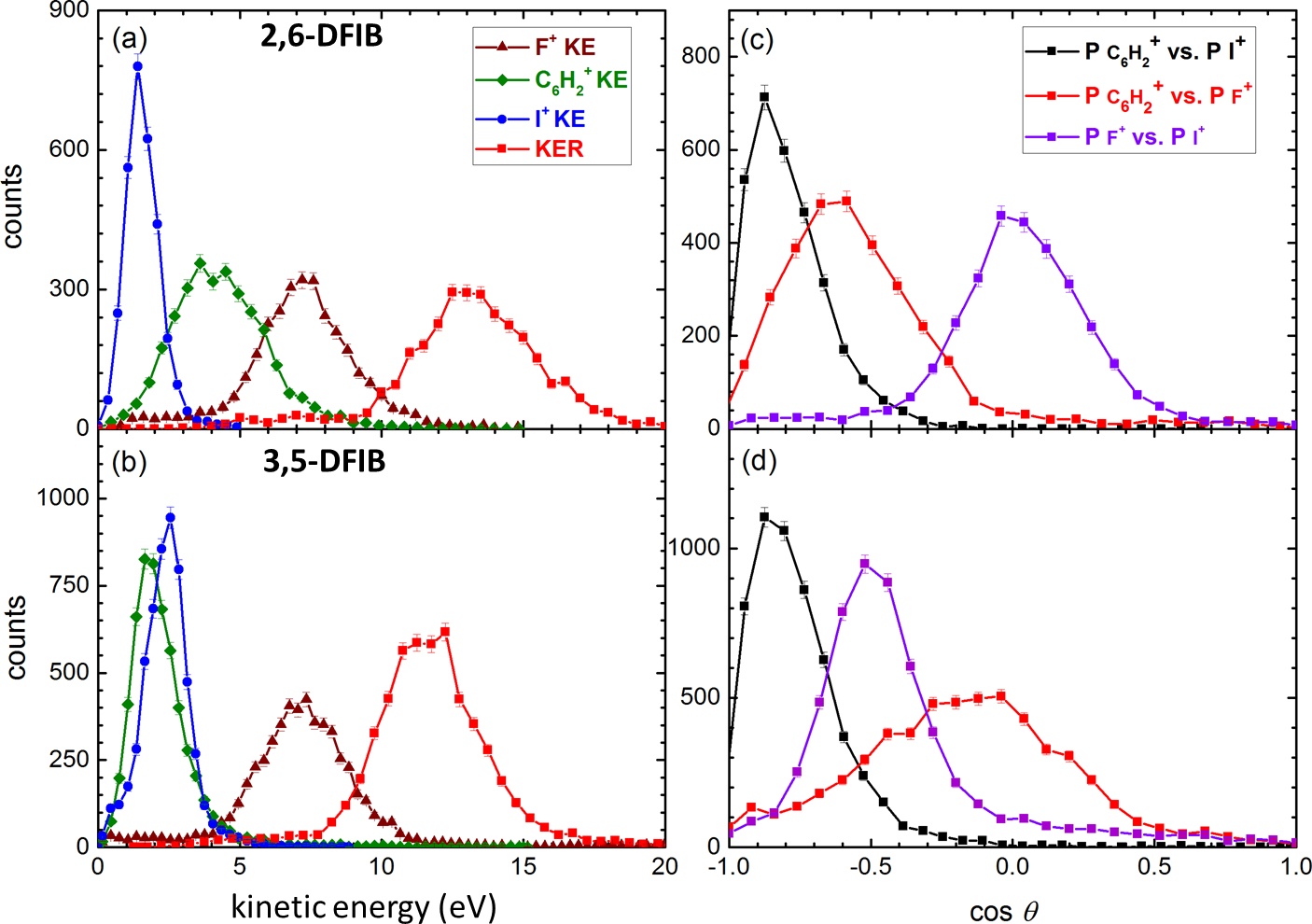}
  \caption{Total kinetic energy release, kinetic energies of the individual ionic fragments, and angle between the momentum vectors of the F$^+$, I$^+$, and C$_6$H$_2$$^{+}$ fragments for the F$^{+}$ + C$_6$H$_2$$^{+}$ + I$^{+}$ fragmentation channel.}
  \label{fgr:26_35_KE_Fp_C6H2p_Ip}
\end{figure}

\begin{figure}[ht]
\centering
  \includegraphics[height=6.5cm]{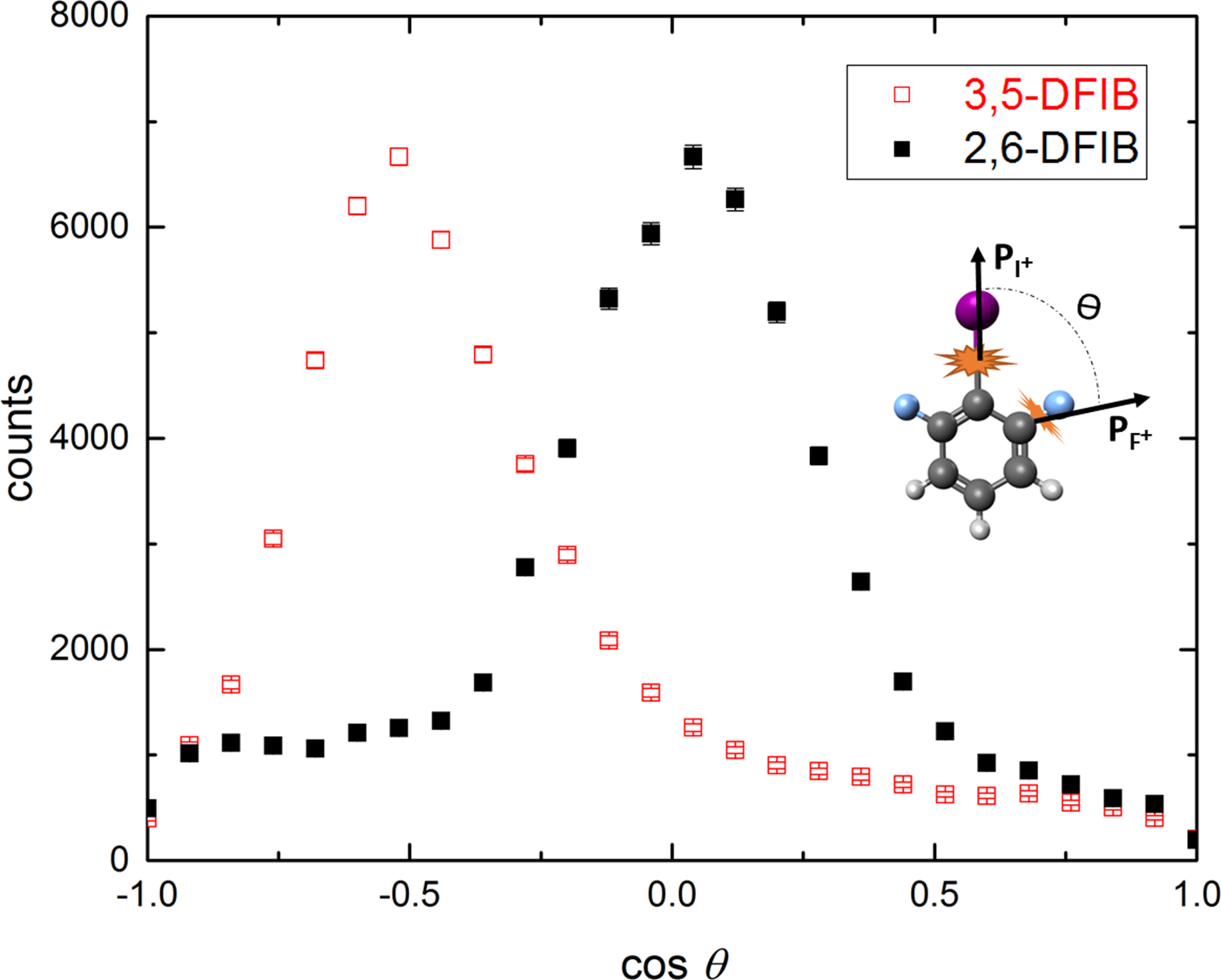}
  \caption{Momentum vector correlation between F$^+$ and I$^+$ fragments in 2,6-DFIB (black solid symbols) and 3,5-DFIB (red open symbols) for fragmentation channels where F$^+$ is the first and I$^+$ is the third detected fragment.}
  \label{fgr:FpIpvector}
\end{figure}
%
Finally, we investigate how general the above findings are for other channels involving F$^{+}$ production. Fig.~\ref{fgr:26_35_KE_Fp_C6H2p_Ip} shows the kinetic energy and momentum angle distributions for the strongest three-body fragmentation channel containing F$^{+}$, namely F$^{+}$ + C$_6$H$_2^{+}$ + I$^{+}$, where the missing fluorine and hydrogen atoms are emitted as one or two neutral fragment(s). For 3,5-DFIB, the distributions are very similar to the F$^{+}$ + C$_6$HF$^{+}$ + I$^{+}$ fragmentation channel, with the angular distributions for F$^{+}$ + C$_6$H$_2^{+}$ + I$^{+}$ being slightly broader. For 2,6-DFIB, both kinetic energy and angular distributions are significantly broadened, while the peak positions are still close the the former case. Furthermore, Fig.~\ref{fgr:FpIpvector} shows the I$^{+}$-F$^{+}$ angle for all events where F$^{+}$ is detected as the first fragment and I$^{+}$ as the last fragment, thus integrating over all possible partner fragments. Again, the I$^{+}$-F$^{+}$ angles are very similar to the F$^{+}$ + C$_6$HF$^{+}$ + I$^{+}$ fragmentation channel discussed above, suggesting that the sequential breakup with a delay of approximately 400 fs is common to all triply charged final states that involve F$^{+}$ production in 2,6-DFIB, while a concerted fragmentation is well suited to describe the breakup in 3,5-DFIB.

\section{Conclusions}
\label{sec:conclusions}
We have presented a detailed study of the photoionization and fragmentation dynamics of inner-shell ionized 2,6- and 3,5-DFIB isomers using coincident electron-ion moment imaging. Our results demonstrate that the coincident electron-ion momentum imaging technique is a powerful method to study even such rather complex molecules consisting of twelve atoms. Fragment ion kinetic energies and angular correlations contain detailed information on the fragmentation dynamics, which can be interpreted using classical Coulomb explosion models. By comparing these model calculations with the experimental observations, we can distinguish different electronic decay dynamics and fragmentation mechanisms. In particular, we conclude that charges on the di- and tri-cation separate on an ultrafast timescale, and that some fragmentation channels of the tri-cation involve step-wise fragmentation with a delay between the two steps ranging from a few hundred femtoseconds to tens or hundreds of picoseconds or longer. Finally, our experimental observations show that the angle between F$^+$ and I$^+$ fragments in three-body fragmentation channels can be used to identify and separate the 2,6- and 3,5-DFIB isomers. However, such a direct link between the molecular geometry and the fragmentation pattern should not be taken for granted since the Coulomb repulsion between the fragments and the exact fragmentation dynamics can easily betray naive expectations. Nevertheless, our study demonstrates how sensitive coincident (ion) momentum imaging is to the molecular geometry and dynamics, thus making it a very promising technique for time-resolved experiments, even on polyatomic targets containing ten or more atoms per molecule.

\section{Acknowledgements}
This work is supported by the Chemical Sciences, Geosciences, and Biosciences Division, Office of Basic Energy Sciences, Office of Science, U.S. Department of Energy, Grant No. DE-FG02-86ER13491 (Kansas group) and DE-SC0012376 (U Conn group). D.R. also acknowledges support through the Helmholtz Young Investigator program for the DESY group. We thank the staff of the Advanced Light Source for their hospitality and their help during the beamtime.

\bibliography{rsc}

\end{document}